\documentclass[a4paper]{aa}
\usepackage{lscape}
\usepackage{graphicx} 
\usepackage{latexsym,amsmath,amssymb}
\usepackage{natbib}
\bibpunct{(}{)}{;}{a}{}{,}

\def \chandra {\it Chandra}

\def \swift {\it Swift}

\def \hcm {\hbox {\ifmmode $ atom cm$^{-2}\else atom cm$^{-2}$\fi}}
\def \arcmin {\hbox{$^\prime$}}
\def \arcsec {\hbox{$^{\prime\prime}$}}

\def \chisq {$\chi ^{2}$}

\def \approxgt{\mathrel{\hbox{\rlap{\lower.55ex \hbox {$\sim$}}
        \kern-.3em \raise.4ex \hbox{$>$}}}}
\def \approxlt{\mathrel{\hbox{\rlap{\lower.55ex \hbox {$\sim$}}
        \kern-.3em \raise.4ex \hbox{$<$}}}}

\def \srctwo {4U\,1820$-$30}
\def \srcone {4U\,1728$-$34}
\def \aql {Aql\,X$-$1}

\begin{document}

\title{ALMA observations of \srcone\ and \srctwo: first detection of neutron star X-ray binaries at 300 GHz}

\author{M. D{\'i}az Trigo\inst{1} \and S. Migliari\inst{2,3} \and J.~C.~A. Miller-Jones\inst{4} \and F. Rahoui\inst{1,5} \and D.~M. Russell\inst{6} \and V. Tudor\inst{4}
}
\institute{
ESO, Karl-Schwarzschild-Strasse 2, D-85748 Garching bei M\"unchen, Germany
              \and
              XMM-Newton Science Operations Centre, ESAC/ESA, Camino Bajo del Castillo s/n, Urb. Villafranca del Castillo, 28691 Villanueva de la Ca\~nada, Madrid, Spain
\and
Department of Quantum Physics and Astrophysics \& Institute of Cosmos Sciences, University of Barcelona, Mart\'i i Franqu\`es 1, 08028 Barcelona, Spain
\and
              International Centre for Radio Astronomy Research, Curtin University, GPO Box U1987, Perth,
WA 6845, Australia
\and
Department of Astronomy, Harvard university, 60 Garden street, Cambridge, MA 02138, USA
\and
New York University Abu Dhabi, PO Box 129188, Abu Dhabi, UAE
}

\date{Received ; Accepted:}

\authorrunning{D{\'i}az Trigo et al.}

\titlerunning{ }

\abstract{We report on the first observations of neutron star low-mass X-ray binaries with the Atacama Large Millimeter/submillimeter Array (ALMA) at $\sim$300~GHz. Quasi-simultaneous observations of \srcone\ and \srctwo\ were performed at radio (ATCA), infrared (VLT) and X-ray (Swift) frequencies, spanning more than eight decades in frequency coverage. Both sources are detected at high significance with ALMA. The spectral energy distribution of \srcone\ is consistent with synchrotron emission from a jet with a break from optically thick to optically thin emission at 1.3--11.0$\times$10$^{13}$~Hz. This is the third time a jet spectral break has been reported for a neutron star X-ray binary. The radio to mm spectral energy distribution of \srctwo\ has significant detections at 5 and 300~GHz. This confirms the presence of radio emission during a soft state for this neutron star and represents the first detection of mm emission during such a state, unambiguously pointing to the presence of a jet. We also report on three additional unrelated sources - showing mm emission - in the ALMA fields of view of \srcone\ and \srctwo.
  
} \maketitle

\section{Introduction}
\label{sect:intro}

Relativistic jets are a common phenomenon associated with accretion onto compact objects. They are observed in X-ray binaries (XRBs) and supermassive black holes and are thought to be a fundamental ingredient of Gamma-Ray Bursts, the most powerful transient events in the Universe. The study of jets in XRBs provides us with the distinctive advantage of witnessing their appearance and disappearance on human timescales (days to weeks). Moreover, since XRB compact objects can be either a neutron star (NS) or a black hole (BH), comparison studies allow us to isolate the role played by the BH event horizon or by the NS surface or magnetic field in powering the jets. 

To date, most of the studies of XRB jets have been performed on BHs. Based on the results, a paradigm has emerged according to which the presence/absence of a compact or transient jet is related to the  accretion flow geometry during a transient outburst \citep{fender04mnras}. In short, during the rising phase of the outburst the steady jet known to be associated with the canonical ``low-hard state" (LHS) persists while the X-ray spectrum initially softens. During the transition to the ``high-soft state" the steady jet emission is quenched \citep[see e.g.][]{russell11apj, 1743:coriat11mnras}. In some sources discrete ejecta are launched \citep[although not in every transition -- ][]{1659:paragi13mnras} and their radio emission can be spatially resolved \citep[e.g.][]{1915:mirabel94nature,1655:tingay95nature}. When the BH transitions back to the LHS at a lower luminosity, the compact jet is reinstated \citep{kalemci13apj}.

In this paper, we are interested in mapping the jet properties of NS XRBs to their ``hard'' and ``soft'' states (we note that for NSs these states are often described in the literature as ``island'' and ``banana'' states, respectively, but we use the former nomenclature hereafter to ease the comparison with BHs). For this, we obtained multi-wavelength observations of two low-magnetic field NSs from the ``atoll'' class \citep{hasinger89aa}: \srcone\ and \srctwo. These are sources that resemble BH XRBs in their spectral and timing properties, as opposed to the higher luminosity ``Z'' class sources that persistently emit close to or above the Eddington luminosity. Previous studies of the ``atoll'' class show a somewhat contradictory picture. The systematic study performed by \citet{migliari06mnras} of the disc-jet coupling in persistent NS XRBs showed that there could be two distinct differences between NSs and BHs. First, NSs may not all completely suppress their jet during the soft state, as radio emission from two NS persistent sources was detected during that state \citep{1820:migliari04mnras}. Second, the radio emission at the top end of the hard state seems to be a factor 30 lower than for their BH counterparts at similar X-ray flux levels \citep{1728:migliari03mnras}. In contrast, \citet{aql:miller-jones10apj} studied the radio and X-ray emission of the transient NS \aql\ during its 2009 outburst and found the radio emission to be consistent with being triggered at state transitions, both from the hard to the soft state and vice versa, just as in BHs. Quenching of the radio emission in the soft state above $\sim$\,10$\%$ of the Eddington luminosity, in agreement with what is found for BHs, was also observed \citep[see also][for quenching of the radio emission in the NS ``atoll'' source GX9+9]{migliari11iaus}. 

For our purpose of comparing jet properties of NSs and BHs in different accretion states, a key ingredient is the measurement of the jet spectral break from optically thick to optically thin synchrotron emission. The frequency of this break together with the frequency of the cooling break expected at higher energies \citep{sari98apj} determine the total radiative power of the jet. The cooling break has only been detected once in XRBs \citep{1836:russell14mnras}. However, the spectral break from optically thick to optically thin emission has recently been extensively studied and shown to be driven primarily by the changing structure of the accretion flow rather than by the mass and spin of the BH or its luminosity \citep{1659:horst13mnras,gx339:corbel13bmnras,1836:russell14mnras,koljonen15apj}. Moreover, since it is observed to move from $\sim$10$^{14}$~Hz down to $\sim$10$^{11}$~Hz as the X-ray spectrum softens, observations around 10$^{11}$~Hz are crucial to compare the jet power during hard and soft states. 

\srcone\ is a low-mass X-ray binary (LMXB) at a distance of 5.2 kpc \citep[as derived from type I X-ray bursts,][]{galloway08apj}. \citet{hasinger89aa} classified it as an atoll-type X-ray binary based on its spectral and timing properties. \citet{1728:marti98aa} detected the radio counterpart of \srcone\ with the Very Large Array (VLA) at 4.86 GHz with a variable flux density ranging between 0.3 and 0.6~mJy, and a J and K-band infrared source (J=19.6 and K=15.1) within one arcsecond of the radio source. \citet{1728:migliari03mnras} observed \srcone\ in 2000-2001 simultaneously with the VLA (radio) and RXTE (in X-rays). They investigated the connection between the radio flux density and X-ray flux and found that it is qualitatively similar to that found for BHs in the LHS, in that the radio and X-ray fluxes positively correlate (except in one case that might be associated to a soft state, see their Fig.~3 and Sect.~5). They also reported the highest radio flux densities (up to 0.6 mJy with variations of 0.3 mJy) in observations of transitional states between the hard state and the soft state and lower, more stable, flux densities of $\sim$0.09-0.16~mJy during observations in the hard state. 

\srctwo\ is an ultra-compact LMXB located in the globular cluster NGC~6624 \citep{giacconi74apj}. The distance derived from type I X-ray bursts is 6.4~kpc \citep{galloway08apj} and that derived from optical observations is 7.6\,$\pm$\,0.4~kpc \citep{1820:heasley00aj}. A radio source at the position of \srctwo\ was first detected in 1.4~GHz VLA observations, with a flux density of 2.44 mJy \citep{1820:geldzahler83apj}. However, the identification of this radio source as the radio counterpart of the LMXB was controversial due to the proximity of the radio pulsar PSR B1820--30A \citep{1820:johnston93aa}. \citet{1820:migliari04mnras} observed \srctwo\ simultaneously with the VLA and RXTE seven times in July and August 2002. They identified the state of the source as being soft based on X-ray spectral and timing characteristics, and detected an average radio flux density of 0.13\,$\pm$\,0.04~mJy and 0.08\,$\pm$\,0.02~mJy at 4.86 and 8.46 GHz, respectively, at J2000 RA 18:23:40.4820 $\pm$ 0.0088, Dec. -30:21:40.12$\pm$0.166. 
They compiled all the previous radio observations performed at $\sim$1.5, 5 and 9~GHz (see their Fig.~4) and compared the spectral indices derived below and above 5~GHz.  
They concluded that the X-ray binary should dominate the emission above 1.5 GHz because of the steep spectral index from the radio pulsar 
(-3.7: \citealt{biggs94mnras}; -2.7\,$\pm$\,0.9: \citealt{toscano98apj}), and the high variability of the radio source \citep{1820:fruchter90apjl,1820:johnston93aa}. They derived a spectral index  of -\,0.48$\pm$0.62 between 5 and 8.5~GHz, consistent with either an optically thick or thin spectrum, as well as with the index derived from previous dual-frequency observations at 1.5 and 5.5~GHz \citep{1820:fruchter00apj}. 

We specifically chose \srcone\ and \srctwo\ for our exploratory study because they are persistently accreting, such that scheduling their observations is easier, and because they are known to spend most of their lives in the hard \citep{hasinger89aa} and soft \citep{1820:titarchuk13apj} states, respectively. We performed quasi-simultaneous observations at radio, millimeter (mm), infrared (IR) and X-ray frequencies, taking advantage of the largely improved sensitivity of the Atacama Large Millimeter/submillimeter Array (ALMA) to study these two sources for the first time at frequencies of $\sim$300~GHz. In this paper, we report on the results of this campaign. 

\section{Observations and data analysis}
\label{sec:observations}

We observed \srcone\ and \srctwo\ in July 2014 at radio, mm, IR and X-ray frequencies with the Australia Telescope Compact Array (ATCA), ALMA, the Very Large Telescope (VLT) and $\swift$ \citep{swift:gehrels04apj}. Table~\ref{tab:obslog} shows a log of the observations.

\begin{table*}
\begin{center}
\caption[]{Observation log. The exposure time excludes time spent on calibrator sources.}
\begin{tabular}{ccllll}
\hline \noalign {\smallskip}
Source & MJD & Observatory & Obs ID & Exposure (ks) & Flux \\
\hline \noalign {\smallskip}
\srcone\  & 56860 & ATCA 5.5 GHz & C3010 & 6.9 & 0.424\,$\pm$\,0.022~mJy beam$^{-1}$ \\
  & 56860 & ATCA 9.0 GHz & C3010 & 6.9 & 0.587\,$\pm$\,0.038~mJy beam$^{-1}$ \\
  & 56857 & ALMA 302 GHz &  2013.1.01013.S & 2.8 &1.40\,$\pm$\,0.02~mJy beam$^{-1}$ \\ 
  & 56859 & VLT/Hawk-I K$_S$ & & 0.060 & 14.48\,$\pm$\,0.02~magnitudes \\
    & 56859 & VLT/Hawk-I J & & 0.012 & 17.49\,$\pm$\,0.03~magnitudes \\
  & 56856 & $\swift/BAT$& -- & -- & 0.014\,$\pm$\,0.003 counts cm$^{-2}$ s$^{-1}$ \\
  & 56860 & $\swift/BAT$& -- & -- & 0.005\,$\pm$\,0.001 counts cm$^{-2}$ s$^{-1}$ \\
& & & & \\
\srctwo\  & 56855 & ATCA 5.5 GHz & C3010 & 10.8 & 0.236\,$\pm$\,0.027~mJy beam$^{-1}$  \\
& 56855 & ATCA 9.0 GHz & C3010  & 10.8 &  $<$0.200~mJy beam$^{-1}$ (see text) \\
  & 56845 & ALMA 302 GHz & 2013.1.01013.S & 2.9 & 0.40\,$\pm$\,0.02~mJy beam$^{-1}$ \\
  & 56841 & $\swift/XRT$ & 00080210020 & 0.6 & 7.52 ($\pm$0.09)\,$\times$\,10$^{-9}$ erg cm$^{-2}$ s$^{-1}$ \\
  & 56857 & $\swift/XRT$ & 00035341004 & 1 & 7.77 ($\pm$0.05)\,$\times$\,10$^{-9}$ erg cm$^{-2}$ s$^{-1}$ \\
  & 56837 & $\swift/BAT$ & -- & -- & 0.012\,$\pm$\,0.001~counts cm$^{-2}$ s$^{-1}$ \\
  & 56845 & $\swift/BAT$ & -- & -- & 0.009\,$\pm$\,0.003~counts cm$^{-2}$ s$^{-1}$ \\
  & 56855 & $\swift/BAT$ & -- & -- & 0.009\,$\pm$\,0.002~counts cm$^{-2}$ s$^{-1}$ \\
  & 56856 & $\swift/BAT$ & -- & -- & 0.016\,$\pm$\,0.002~counts cm$^{-2}$ s$^{-1}$ \\
\noalign {\smallskip} \hline 
\label{tab:obslog}
\end{tabular}
\end{center} 
\end{table*}

\subsection{ATCA}
\label{sec:ATCA}
\subsubsection{\srcone}

We observed \srcone\ with the Australia Telescope Compact Array (ATCA) on the 22nd of July 2014, from 09:00 to 11:00 UTC.  The array was in the compact H75 configuration, whereby the maximum baseline between the five inner antennas was 89~m, with the sixth antenna located 4.4~km away. We used the Compact Array Broadband Backend \citep{Wilson2011} to observe simultaneously at 5.5 and 9.0\,GHz, using 2048~MHz of bandwidth at each frequency.  We used B1934-638 as our bandpass calibrator and to set the amplitude scale, and the more nearby, compact source, B1714-336 as our phase calibrator.  We reduced the data according to standard procedures in the Multichannel Image Reconstruction, Image Analysis and Display software package \citep[MIRIAD v1.5;][]{Sault1995}, and then wrote them out to uvfits format for subsequent imaging in the Common Astronomy Software Application \citep[CASA v4.2.1;][]{McMullin2007}. 
However, owing to the short duration of the observation, the {\it uv}-coverage was poor, and CASA was unable to handle the highly elongated beam shape when all six antennas were used.  We therefore performed the final imaging and image-plane fitting using the Astronomical Image Processing System \citep[AIPS v31DEC12;][]{Greisen2003}, which was better able to handle the elongated beam shape.

Imaging with all six baselines antennas picked up significant diffuse emission from the field.  Pure uniform weighting using all six antennas was able to detect compact emission from the location of \srcone\, but it was still contaminated by a varying background level from the diffuse emission to which the short baselines were sensitive.  We therefore restricted our imaging to include only the longer baselines to antenna 6, thus enabling us to reliably constrain the compact emission from the X-ray binary.  We detected the target source at flux densities of $424\pm22$ and $587\pm38$\,$\mu$Jy\,beam$^{-1}$ at 5.5 and 9.0\,GHz, respectively.

\subsubsection{\srctwo}
We observed \srctwo\ with the ATCA on the 17th of July 2014, from 10:00 to 15:00 UTC.  The array was once again in its compact H75 configuration.  However, two antennas (including the distant antenna 6) were offline during our observations, leaving us with an array of 6 short baselines. We used an identical frequency setup to that employed for \srcone. We used B1934-638 as our bandpass calibrator and to set the amplitude scale, and B1817-254 as our secondary calibrator, to determine the instrumental complex gains that could be interpolated onto the target field. Once again, data reduction was performed according to standard procedures within MIRIAD, before exporting the data for imaging in CASA.  

\srctwo\ is located in the globular cluster NGC\,6624, and is only 3.5\arcmin\ from a bright (14~mJy) background source \citep{Knapp1996}. The presence of this source at the 64\% and 34\% response points of the primary beam at 5.5 and 9.0~GHz, respectively, meant that slight antenna pointing errors led to amplitude errors in the data that could not be effectively corrected, since there was not enough flux in the image to self-calibrate on a timescale shorter than 10~min.  Deconvolution errors both from this bright source and from unresolved, fainter sources in the field, meant that the noise level in the vicinity of the target was higher than expected.  We therefore used a model made from deep images of NGC\,6624 at the same observing frequencies (Tudor et al., in prep.), taken with the ATCA in its extended 6~km configuration, to subtract out (in the {\it uv}-plane) all emission from compact sources in the field, with the exception of \srctwo. Imaging the residual data at 5.5~GHz showed that this had both reduced the brightness of the confusing source, and removed additional confusing emission from faint, compact sources in the field, to the point that we significantly detected a source at the location of \srctwo, with a flux density of $236\pm27$~$\mu$Jy\,beam$^{-1}$.  At 9.0\,GHz however, we did not detect any significant emission at the location of the X-ray binary, to a $3\sigma$ upper limit of 135~$\mu$Jy\,beam$^{-1}$. To further test the robustness of this upper limit we added simulated point sources to the visibility data at a range of position angles at the same angular distance from \srctwo\ as the confusing source and imaged the resulting data with a Briggs robust weighting of zero (this weighting gave the best noise level far away from the confusing source). On varying the flux density of the inserted sources in steps of 50 $\mu$Jy beam$^{-1}$, we found that the faintest source that could be reliably detected at all position angles was 200~$\mu$Jy beam$^{-1}$. We consider this upper limit as more realistic and use it hereafter.

\subsection{ALMA}

ALMA observed \srcone\ on the 19th of July 2014 from 01:05 to 03:32 UTC and \srctwo\ on the 7th of July 2014 from 04:27 to 07:01 UTC. Both observations were set up to use four spectral windows with a bandwidth of 2 GHz each and centred at 295.987, 297.924, 306.002 and 308.002 GHz. The spectral resolution for each spectral window was 31.2 MHz or $\sim$31~km s$^{-1}$. The observations were performed using 31 and 33 12-m antennas for \srcone\ and \srctwo, respectively, with a maximum baseline length of 650~m. For \srcone, the bandpass, flux, and phase calibrators were J1733$-$1304, Titan, and J1744$-$3116, respectively. For \srctwo, all calibrations were performed with J1924$-$2914. 

\begin{figure*}[ht]
\hspace{0cm}\includegraphics[angle=0.0,width=0.33\textheight]{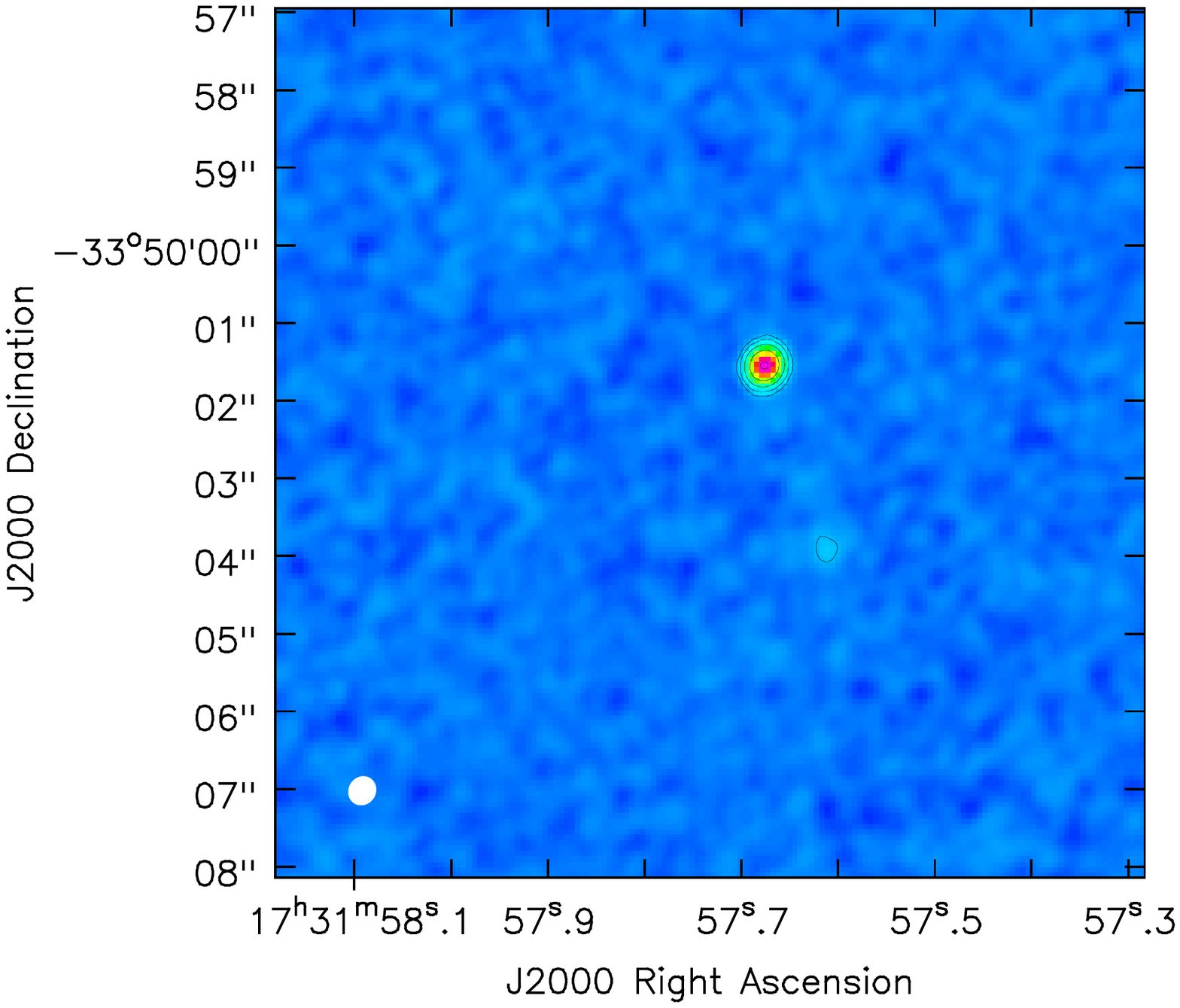}
\hspace{0cm}\includegraphics[angle=0.0,width=0.33\textheight]{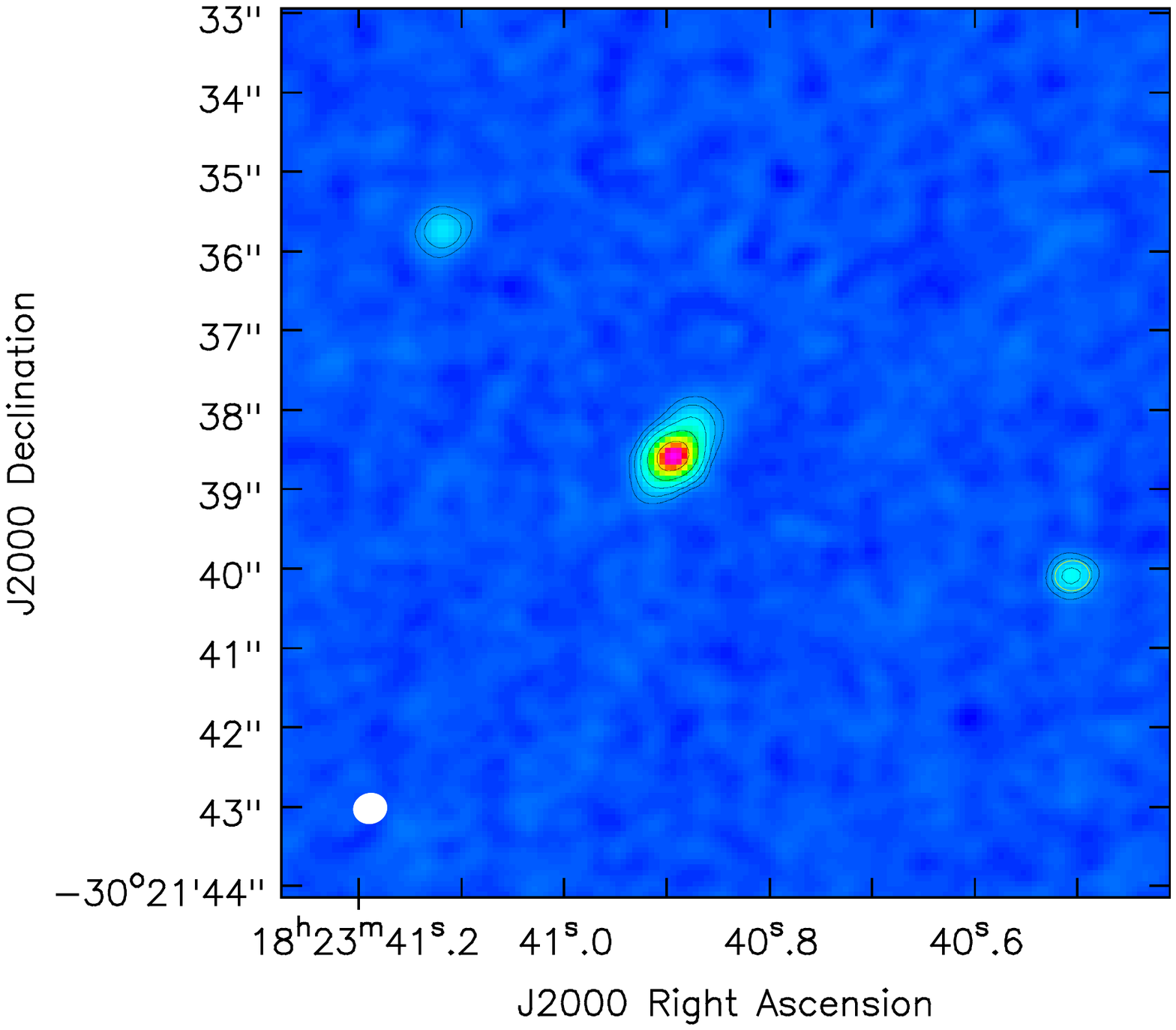}
\vspace{-4cm}
\caption{ ALMA images of the averaged data of \srcone\ (left) and \srctwo\ (right) at 301.99~GHz. The images are cutouts of the full field of view (half power beam width of 17\arcsec) centred on the detected sources. {\it Left panel}: \srcone\ is the brightest source in the ALMA field-of-view with a flux density of 1.40~mJy~beam$^{-1}$. A second, weaker source is detected to the southwest of \srcone\ with a flux density of 0.10~mJy~beam$^{-1}$. {\it Right panel}: \srctwo\ is the source at the bottom-right of the image, with a flux density of 0.40~mJy~beam$^{-1}$. Two additional sources are visible to the northeast, one of which is marginally extended (see Table~\ref{tab:sources}). 
Contours are overlaid in the zoomed images at levels of $\pm (2)^n$ $\sigma$ significance with respect to a noise of $\sigma=20$~$\mu$Jy beam$^{-1}$, where $n$ = 2, 3, 4, 5...15. The synthesized beam is shown at the lower-left corner of each image.}
\label{fig:image1728}
\end{figure*}

We used the ALMA calibrated products to re-generate images following standard procedures. CASA was used for image generation and analysis. The reconstructed images at the positions of \srcone\ and \srctwo, extracted with uniform weighting and using a central frequency of 301.99~GHz, are shown in Fig.~\ref{fig:image1728}. The images were self-calibrated on a timescale of 1 hour. The flux densities and uncertainties were determined after correcting the maps for primary beam attenuation.

For \srcone\ the synthesized beam size was 0.33\arcsec$\times$0.29\arcsec. The peak of radio emission  is centred at (J2000) 17:31:57.6757\,$\pm$\,0.0003 -33:50:01.544\,$\pm$\,0.005 
and the peak flux density is 1.40\,$\pm$\,0.02~mJy beam$^{-1}$ (the reported flux density for all sources in this section corresponds to the peak flux from the elliptical Gaussian fitted to the source emission in the image plane and the error corresponds to the rms of the image measured far away from the source position). We detect a second, weak, source to the southwest of \srcone, with a flux density of 0.10\,$\pm$\,0.02~mJy beam$^{-1}$ (we name all the new sources according to the ALMA new source nomenclature at the ALMA Users' Policies, see Table~\ref{tab:sources}).

For \srctwo\ the synthesized beam size was 0.34\arcsec$\times$0.32\arcsec. In the ALMA field-of-view we detect three sources (see Table~\ref{tab:sources}), one of which, ALMA J182340.892--302138.56, shows extended emission. We identify \srctwo\ with the source at the centre of the ALMA image at a position of (J2000) 18:23:40.5050\,$\pm$\,0.0004 -30:21:40.089$\pm$0.005 and with a peak flux density of 0.40\,$\pm$\,0.02 mJy beam$^{-1}$. The possible origin of the other two sources is discussed in Sect.~\ref{sec:discussion}.

\begin{table*}
\begin{center}
\caption[]{Positions and flux densities of sources detected in the ALMA images. All the sources are detected for the first time at mm frequencies. ALMA J182340.8936--302138.571 is marginally extended with respect to the ALMA synthesized beam and has an integrated flux of 3.71\,$\pm$0.13 mJy.}
\begin{tabular}{lllc}
\hline \noalign {\smallskip}
Source & RA (J2000) & Dec (J2000) & Flux [mJy~beam$^{-1}$] \\
\hline \noalign {\smallskip}
\srcone\ & 17:31:57.6757\,$\pm$\,0.0003 & -33:50:01.544\,$\pm$\,0.005 & 1.40\,$\pm$\,0.02 \\
ALMA J173157.613--335003.88 & 17:31:57.613\,$\pm$\,0.002 & -33:50:03.88\,$\pm$\,0.03 & 0.10\,$\pm$\,0.02 \\
\srctwo\ & 18:23:40.5050\,$\pm$\,0.0004 & -30:21:40.089\,$\pm$\,0.005 & 0.40\,$\pm$\,0.02 \\
ALMA J182340.8936--302138.571 & 18:23:40.8936\,$\pm$\,0.0003 & -30:21:38.571\,$\pm$\,0.005 & 2.26\,$\pm$\,0.02\\
ALMA J182341.1181--302135.733 & 18:23:41.1181\,$\pm$\,0.0007 & -30:21:35.733\,$\pm$\,0.009 & 0.50\,$\pm$\,0.02 \\
\noalign {\smallskip} \hline 
\label{tab:sources}
\end{tabular}
\end{center} 
\end{table*}

\subsection{VLT}

On the 22nd of July 2014, \srcone\ was observed in the J and Ks bands with the High Acuity Wide field K-band Imager \citep[HAWK-I,][]{2004Pirard, 2006Casali, 2008Kissler, 2011Siebenmorgen}, mounted on UT4 at the Very Large Telescope (VLT) at Cerro Paranal. The individual exposure time was set to 2~s in both filters, and standard 140\arcsec\ dithering was used for sky background subtraction, leading to total exposure times of 12~s and 60~s in the $J$ and $Ks$ bands, respectively. Seeing at 500~nm as measured on the images was in the range 0\farcs88--1\farcs12\ and the airmass was lower than 1.02.

The data were reduced with the dedicated HAWK-I pipeline (v1.8.18) provided by ESO and implemented in the EsoRex package. The procedure we followed is standard for wide-field infrared mosaic imaging and includes dark and bias subtraction, bad pixel removal, flatfielding, two-pass background subtraction, distortion correction, astrometric calibration, and mosaic projection. Zero point magnitudes were then assessed using the photometric standard star P565-C, which was observed under the same conditions, and we derived $Z_{\rm PJ}\,=\,26.48\,\pm\,0.01$ and $Z_{\rm PKs}\,=\,25.81\,\pm\,0.01$. Using aperture photometry, we finally measured the magnitudes of \srcone as $m_{\rm J}\,=\,17.49\,\pm\,0.03$ and $m_{\rm Ks}\,=\,14.48\,\pm\,0.02$, respectively.

\subsection{Swift}
 
The $\swift$ X-ray telescope \citep[XRT, ][]{swift:burrows05ssr} observed \srctwo\ on the 3rd and the 19th of July 2014 in Window Timing Mode for 564 and 960 seconds, respectively. For the analysis, we used HEASOFT v6.19 and the most recent calibration files and followed the standard procedures (see XRT data analysis threads at http:$\slash\slash$www.swift.ac.uk$\slash$analysis$\slash$xrt$\slash$index.php). We re-created the event level 2 files using as source position (J2000) 18:23:40.57 -30:21:40.6. We extracted image, light curve and spectrum using as source region a circle of 30 pixels centred on the source position and as background a circle of radius 30 pixels far from the sources. We note that given the high count rate of the source, the background subtraction had no effect and was therefore not performed.  The light curve was extracted in the 0.3--10 keV energy range. As suggested in the $\swift/XRT$ data analysis threads, when extracting the spectrum we filtered only grade 0 because the source is heavily absorbed \citep[hydrogen equivalent column density: N$_{H}$ $> 10^{21-22}$ cm$^{-2}$,][]{nh:dickey90araa}. 

We grouped the bins of the spectrum with 20 counts minimum per bin in order to use \chisq\ statistics. We fitted the 0.3--10 keV spectrum adding 3\% systematic error as suggested by Beardmore et al. 2011\footnote[1]{http:$\slash\slash$www.swift.ac.uk$\slash$analysis$\slash$xrt$\slash$files$\slash$SWIFT-XRT-CALDB-09\_v16.pdf}. 

The spectra were fitted in the energy range 0.3--10~keV fixing N$_{H}$ = 0.15$\times$10$^{22}$ cm$^{-2}$ \citep{nh:dickey90araa}. We tried different combinations of absorbed power law, disk blackbody and blackbody models. Three combinations gave a similar fit quality: 1) absorbed disc blackbody and power law, 2)  absorbed blackbody and disc blackbody and 3)  absorbed blackbody and power law. 

We found unabsorbed 2--10~keV fluxes of 7.52\,$\pm$\,0.09 and 7.77\,$\pm$\,0.05 $\times$10$^{-9}$ erg cm$^{-2}$ s$^{-1}$ for the first and second observations, respectively, with any of the three two-component models. We note that for the first observation the value we give is an average of the fluxes obtained with the three different models, while for the second observation all the models give the same flux within uncertainties. 

\section{Results}
\label{sec:results}
\subsection{\srcone}
\label{sec:results1728}

Unfortunately, there were no pointed observations with $\swift/XRT$ within at least one month of the ATCA, ALMA and VLT observations (our trigger was not executed because of other higher priority observations). However, the $\swift/BAT$ all-sky monitor \citep{krimm13apjss} scanned the sky at the position of \srcone\ one day before the ALMA observation and again four days later, on the day of the ATCA observation (see Table~\ref{tab:obslog}). Therefore, 
we determined the state of the source from those two $\swift/BAT$ observations. 

The $\swift/BAT$ 15--50~keV count rates were 0.014\,$\pm$\,0.003 and 0.005\,$\pm$\,0.001~counts cm$^{-2}$ $s^{-1}$ one day before and three days after the ALMA observation, respectively. The corresponding 15-50~keV fluxes are 8.2\,$\times$\,10$^{-10}$ and 2.9\,$\times$\,10$^{-10}$ erg~cm$^{-2}$~s$^{-1}$ assuming a power law of index 2, and 8.7\,$\times$\,10$^{-10}$ and 3.1\,$\times$\,10$^{-10}$ erg~cm$^{-2}$~s$^{-1}$ assuming a power law of index 1. 
Next, we calculated the 15--50~keV fluxes typical of hard and soft states based on previous observations of the source with broad-band X-ray detectors \citep{1728:falanga06aa,1728:tarana11mnras,1728:seifina11apj}. We found values below 10$^{-9}$~erg~cm$^{-2}$~s$^{-1}$ for the majority of soft state observations, while only one hard state observation had a value near this threshold \citep[$\sim$9.8\,$\times$\,10$^{-10}$ erg~cm$^{-2}$~s$^{-1}$, see observation 3 from][]{1728:tarana11mnras}. During the first observation, the 15-50~keV flux does not allow us to discriminate between hard and soft states, but the source must be in a soft state during the second observation. Therefore, given the hard X-ray flux variability and the fact that the ATCA flux reported at 9~GHz (obtained simultaneously to the second  $\swift/BAT$ observation) is similar to fluxes obtained by \citet{1728:migliari03mnras} during transitional states, we conclude that we most likely caught \srcone\ during a transitional state from hard to soft.
If true, the results hereafter should be taken with caution since spectral variations are expected to be strong during such transitional states and our observations at different frequencies are not strictly simultaneous.

\begin{figure}[ht]
\includegraphics[angle=0.0,width=0.35\textheight]{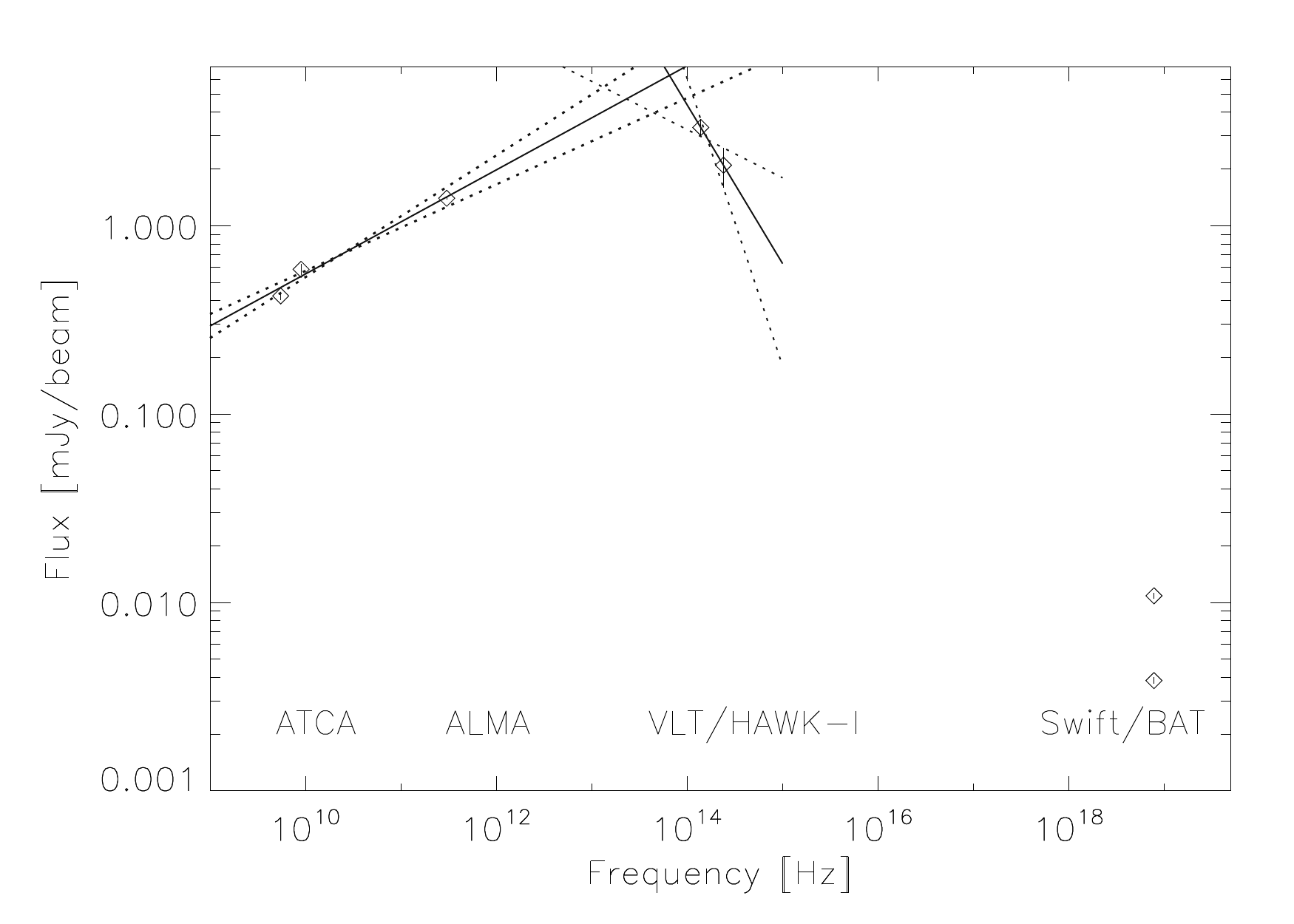}
\caption{Spectral energy distribution for \srcone. We fit the broad-band spectrum with power laws of indices $\alpha$\,=\,0.28\,$\pm$\,0.05 and $\alpha$\,=\,-0.8\,$\pm$\,0.7 below and above $\sim$6.5$\times$10$^{13}$~Hz, respectively. The $\swift/BAT$ points correspond to the flux densities derived in Sect.~\ref{sec:results1728} for observations before and after the ALMA observations (see Table~\ref{tab:obslog}).}
\label{fig:sed1728}
\end{figure}

We next fitted the spectrum from radio to IR wavelengths excluding the point at the highest frequency, since it is apparent that there is a break in the spectral slope near the frequency of the first IR measurement, 1.4$\times$10$^{14}$~Hz (see Fig.~\ref{fig:sed1728}). We obtain an index $\alpha$~=~0.18\,$\pm$\,0.02, where $F_{\nu} \propto \nu^{\alpha}$ and $F_{\nu}$ is the flux at frequency $\nu$ \citep{fender01mnras}. This index is consistent with the spectrum being ``inverted'', the result of optically thick synchrotron emission from a jet. Fitting only the last two points of the Spectral Energy Distribution (SED) we obtain a spectral index of  $\alpha$ = -0.8\,$\pm$\,0.7, indicating optically thin synchrotron emission from the jet. Since the ALMA flux is overestimated with respect to the fitted spectral slope if the break is assumed to be at $\sim$1.4$\times$10$^{14}$~Hz, we next considered the (most likely) case that the break is located below the first VLT/HAWK-I point and fitted the low-frequency/optically-thick spectrum considering only ATCA and ALMA points. In this case we obtain $\alpha$\,=\,0.28\,$\pm$\,0.05 and a spectral break at $\sim$6\,$(\pm$\,5)$\times$10$^{13}$~Hz, from interpolation between the optically thick and optically thin spectra. 

\begin{figure}[ht]
\hspace{-0.5cm}\includegraphics[angle=0.0,width=0.4\textheight]{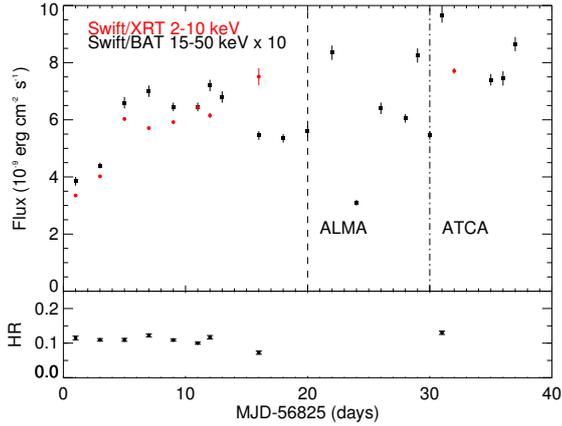}
\vspace{-5cm}
\caption{{\it Top:} $\swift/XRT$ and $\swift/BAT$ fluxes of \srctwo\ before, during and after the ALMA and ATCA observations. {\it Bottom:} Hardness ratio (HR) between the $\swift/BAT$ and $\swift/XRT$ fluxes. Note that the last HR point was calculated between the BAT and the XRT fluxes reported for two consecutive days since after the day of the ALMA observation no measurements were available within the same day. The $\swift/XRT$ fluxes are calculated using a model of two components: a disc blackbody and a power law. The $\swift/BAT$ fluxes are the average between the fluxes obtained assuming a power law of index 1 and a power law of index 2 and the error corresponds to the difference between these fluxes (note that the power-law indices of the fits to the $\swift/XRT$ spectra range between 1 and 2 except for the last observation, for which the index is 0.6\,$\pm$\,0.3, and a fit with two thermal components has been judged as more scientifically meaningful).}
\label{fig:hr1820}
\end{figure}

\subsection{\srctwo}
\label{sec:results1820}

\begin{figure}[ht]
\includegraphics[angle=0.0,width=0.35\textheight]{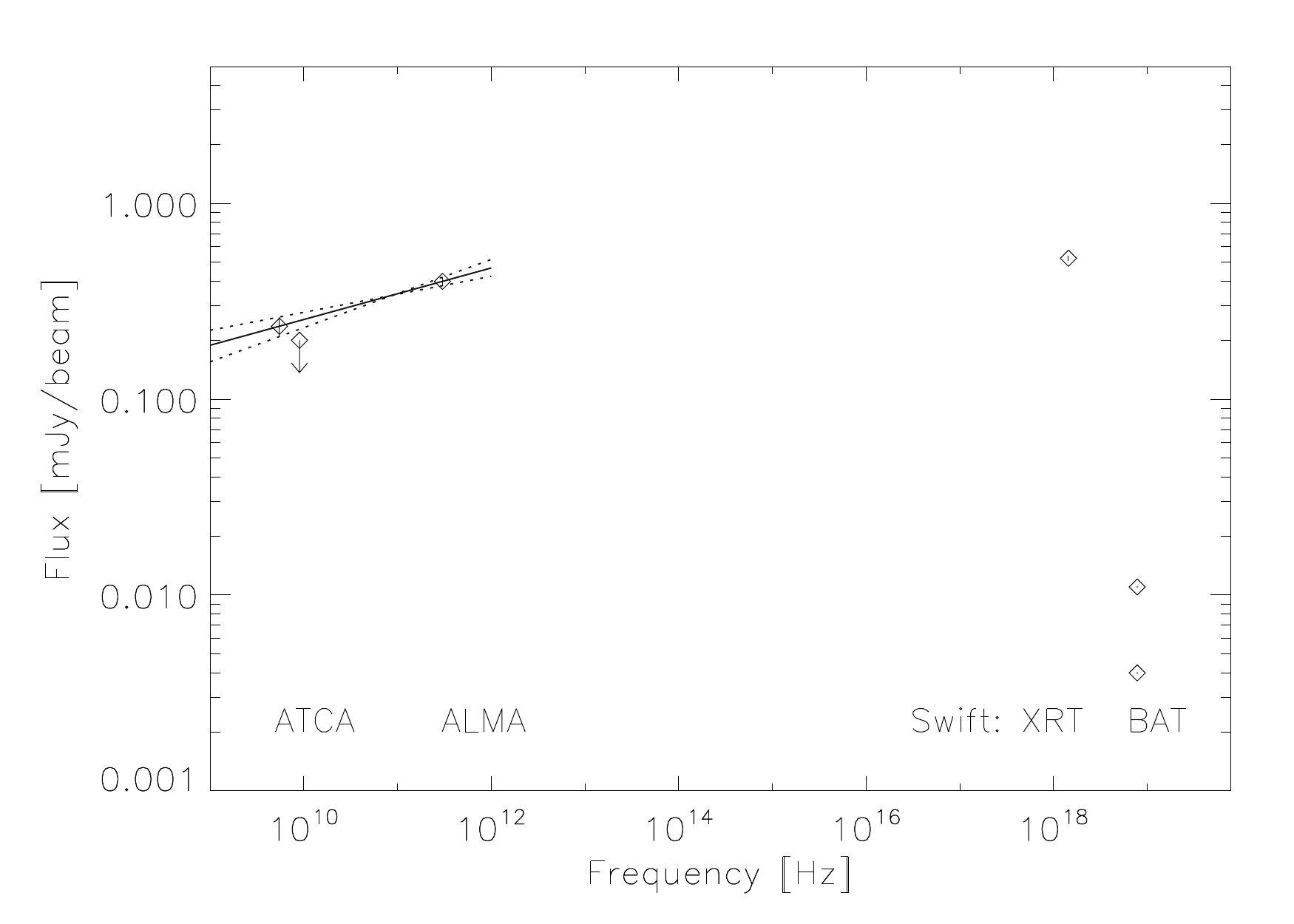}
\caption{Spectral energy distribution for \srctwo. We fit the broad-band spectrum with a power law of index $\alpha$\,=\,0.13\,$\pm$\,0.04. The upper limit at 9~GHz has not been considered to perform the fit. The $\swift/XRT$ point corresponds to the (almost identical) fluxes on the days before and after the ALMA and ATCA observations (see Table~\ref{tab:obslog}). The $\swift/BAT$ points correspond to the flux densities derived in Sect.~\ref{sec:results1820} for observations on the days before and after the ALMA and ATCA observations (see Table~\ref{tab:obslog}).}
\label{fig:sed1820}
\end{figure}

The two $\swift/XRT$ pointed observations were performed four days before and twelve days after the ALMA observation. However, similarly to \srcone, there are $\swift/BAT$ observations of the position of \srctwo\ coincident with the $\swift/XRT$ observations and on the days of the ALMA and ATCA observations that can help us determine the state of the source. 

Following the previous section, we calculated the 15--50~keV fluxes typical of hard and soft states based on previous observations of \srctwo\ with broad-band X-ray detectors \citep{1820:bloser00apj,1820:migliari04mnras,1820:tarana07apj,1820:titarchuk13apj}. Again, we found that values below 10$^{-9}$ erg~cm$^{-2}$~s$^{-1}$ were associated to soft state observations. During the first $\swift/XRT$ observation, the $\swift/BAT$ 15--50 keV flux was 0.012\,$\pm$\,0.001 counts cm$^2$~s$^{-1}$, equivalent to 7.0 (7.4) $\times$10$^{-10}$ erg~cm$^{-2}$~s$^{-1}$ if a power law of index 2 (1) is assumed. This and the ratio of soft to hard flux ($\sim$10) clearly point to a soft state of the source at this time. The ALMA observation took place 4 days later, at which time the hard X-ray flux had decreased by $\sim$25\% indicating that the source was still in a soft state. That flux persisted until the time of the ATCA observation ten days later. Finally, one day after the ATCA observation the hard flux increased again but was still below 10$^{-9}$ erg~cm$^{-2}$~s. At this time, the ratio of the soft to hard flux was $\sim$8. In summary, we conclude that \srctwo\ was in a soft state throughout the observations reported in this paper. This is also supported by the evolution of the X-ray hardness ratio (15--50 keV/2--10~keV fluxes), which remains around $\sim$0.1 before and after the ALMA and ATCA observations (see Fig.~\ref{fig:hr1820}, where the minimum and maximum HR values achieved are 0.073\,$\pm$\,0.005 and 0.125\,$\pm$\,0.005 at days 56841 and 56856, respectively). For comparison, typical HRs during hard states range between $\sim$0.4 and 1.4 \citep[e.g.][]{1820:titarchuk13apj}.

Next, we calculated the spectral slope between radio and mm frequencies with the ATCA and ALMA fluxes. Given that the $\swift/BAT$ flux did not change between the two observations and that the soft state of the source persisted from at least ten days before the ALMA observation and until the ATCA observation, we assumed that the spectral slope can be reliably calculated even if the observations were taken ten days apart. Considering only the two detections at 5.5 and 302~GHz we obtain a spectral slope of 0.13\,$\pm$\,0.04. As discussed in Sect.~\ref{sec:ATCA}, the lack of a detection at 9~GHz could be due to the ATCA data reduction being complicated by the configuration of the array and the field being contaminated by other sources lying in the globular cluster, and \srctwo\ lying only 3.5\arcmin\ from a bright background source. The fitted spectral slope would predict a minimum flux density of 0.228~mJy~beam$^{-1}$ at 9~GHz, not far from our upper limit of 0.200~mJy~beam$^{-1}$. We also consider the possibility that the emission at 5.5~GHz is contaminated by emission from pulsar PSR~B1820--30A \citep{biggs94mnras}. In this case, as discussed by \citet{1820:migliari04mnras} we expect that the pulsar emission is not present at all at 9~GHz due to the steep spectral slope characteristic of the radio pulsar. However, the same spectral slope makes the contribution of the pulsar already negligible at 5.5~GHz \citep[see Fig.~4 at][]{1820:migliari04mnras}, even if the pulsar were scintillating at a level as high as detected by \citet{1820:geldzahler83apj} at 1.5~GHz. Other possibilities are that the spectrum between radio and mm frequencies changed between the ALMA and ATCA observations despite the relatively stable X-ray flux, or that the radio and mm components are unrelated \citep[e.g.][]{a0620:muno06apj}. However, we note that in the latter case, it remains challenging to explain the relatively high ALMA flux with a blackbody component from dust, as used by \citet{a0620:muno06apj} to explain the mid-IR emission. Finally, we cannot exclude the existence of an additional source near the centre of the globular cluster \citep[located at 18:23:40.51, -30:21:39.7 $\pm$\,0.1\arcsec, ][]{1820:goldsbury10aj}, which is not resolved within the ALMA or ATCA beams \citep[see, e.g.][for suggestions of a large number of dark remnants in the central region of the cluster]{1820:peuten14apj}. Such potential source could significantly contribute at the ALMA band but not at the radio or X-ray bands, which would remain dominated by the emission from \srctwo. If we ignore the detection at 5.5~GHz and use instead the upper limit at 9 GHz and the detection at 302~GHz we derive a slope of $>$0.18.

In summary, the most plausible explanation for the non-detection at 9~GHz is the quality of the image at that frequency. However, we cannot exclude variability of the source given the time span between the ATCA and ALMA observations, or the existence of an additional unresolved source at the centre of the globular cluster.

\section{Discussion}
\label{sec:discussion}

\subsection{Radio/X-ray luminosity}

During the hard state of BH accretion outbursts a strong correlation is found between the radio and X-ray luminosities \citep{gx339:corbel03aa, gx339:gallo03mnras}. The form of the correlation, L$_R \propto$ L$_X^b$, has remained unchanged since it was first reported. However, there now appear to be two luminosity tracks (instead of one) represented by b=0.63 and b=0.98, and a transition between the two tracks at $L_X \approx 10^{36}$\,erg\,s$^{-1}$ \citep{1743:coriat11mnras, 1659:jonker12mnras, 1752:ratti12mnras}, or perhaps a large scatter around the original relation \citep{gallo14mnras}. In any case, the radio emission during the hard state is partially self-absorbed, with a flat or inverted spectrum, and has been shown to be the signature of a compact jet \citep{blandford79apj}. 

The radio/X-ray correlation has been studied for NS XRBs by \citet{migliari06mnras}, who found that, similar to BHs, the radio and X-ray luminosities in atoll NSs were also correlated as L$_R \propto$ L$_X^b$. However, b was found to be $\sim$1.4, i.e. atoll NSs were less radio-loud for the same X-ray luminosity. They also found that at least some NSs did not show radio quenching during soft states \citep{1820:migliari04mnras}, in contrast to BHs. 

The latter is confirmed by our observations of \srctwo, which extend the low-frequency emission up to the mm band for the first time. We find a 5.5~GHz flux density and a 9~GHz upper limit that are nearly twice as high as the average flux densities reported by \citet{1820:migliari04mnras} at 4.86~GHz and 8.46~GHz, while our 2--10~keV flux is $\sim$10-40\% lower.  

For \srcone\ in the hard state, \citet{1728:migliari03mnras} reported a flux density of 0.09--0.16 mJy beam$^{-1}$ at 8.46~GHz and a flux of 0.6-0.7\,$\times$\,10$^{-9}$ erg cm$^{-2}$ s$^{-1}$ at 2-10~keV. During transitional states between the hard and the soft state the corresponding radio flux density increases to 0.3-0.6~mJy beam$^{-1}$ and the X-ray flux to 1-2\,$\times$\,10$^{-9}$~erg~cm$^{-2}$~s$^{-1}$. We measured a flux density of 0.6~mJy beam$^{-1}$ at 9~GHz, consistent with the flux density found by \citet{1728:migliari03mnras} in transitional states. The corresponding 2--10~keV flux given by \citet{1728:migliari03mnras} is also consistent for the first X-ray observation and slightly higher for the second observation, assuming that such observations caught the source in a hard and a soft state, respectively. 

\begin{figure}[ht]
\includegraphics[angle=0.0,width=0.35\textheight]{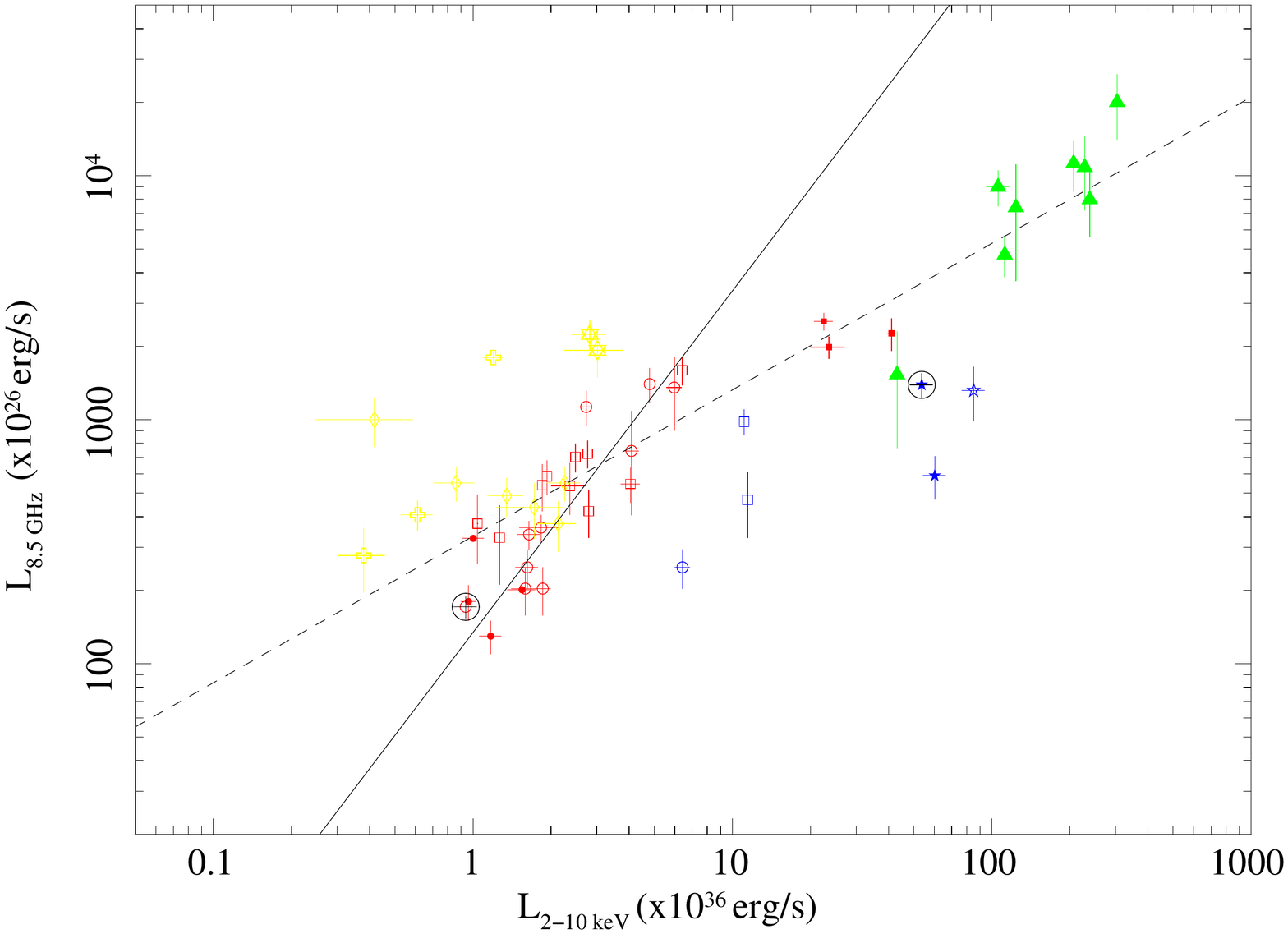}
\caption{Radio (8.5 GHz) luminosity as a function of X-ray (2--10 keV) luminosity of NS XRBs (adapted from \citet{migliari11mnras}). Colours: yellow, accreting milli-second pulsars; red, atoll sources in the hard state or in outburst; blue, atoll sources steadily in the soft state; green, Z sources. Markers: \srcone, open circles; Aql X--1, open squares; MXB 1730--335, filled squares; 4U 0614+091, filled circles; \srctwo, filled stars; Ser X--1, open star. The new observations presented in this paper are marked with black circles. Note that for \srctwo\ we show the extrapolation of the 5~GHz flux to 8.5~GHz assuming an inverted spectrum with a slope of index 0.13, as derived in Sect.~\ref{sec:results1820} instead of the upper limit from the observations reported here.}
\label{fig:radio-x}
\end{figure}

For comparison with previous NS observations, we show in Fig.~\ref{fig:radio-x} the radio (8.5 GHz) luminosity as a function of X-ray (2--10 keV) luminosity of NS XRBs, with the two new points from our 2014 observations. The two new points are roughly consistent with previous measurements of the corresponding sources at similar accretion states.

\subsection{Jet spectral break}

The radiative power of the jet is critically dependent on the position in the spectrum of the break from optically thick to optically thin synchrotron emission \citep[e.g.][]{cygx-1:rahoui11apj} and, at higher frequencies, the cooling break. The latter break is difficult to measure due to the superposition of other emission components at frequencies higher than IR, such as the accretion disc or a hot coronal flow, and has so far only been detected once \citep{1836:russell14mnras}. In contrast, a major effort has been expended in recent years into scheduling multi-wavelength campaigns with the aim of measuring the break at lower frequencies, and obtaining hints that help understand the parameters relevant for jet emission. This effort has been mostly directed to BH XRBs, for which spectral breaks have been reported in nine cases \citep{cygx-1:rahoui11apj, russell13mnras,1836:russell13mnras}. In contrast, NSs have so far been poorly studied due to their faintness in the radio and infrared bands, and a spectral break has been best constrained for 4U~0614+091 \citep{0614:migliari10apj}. For this source, the flat spectrum of index 0.03\,$\pm$\,0.04 was found to break in the range 1--4\,$\times$\,10$^{13}$ Hz to an optically thin power-law synchrotron spectrum with index $\sim$--0.5. Based on the spectral slope of the optically thin part of the spectrum, \citet{0614:migliari10apj} estimated a lower limit to the jet radiative power of $\sim$\,3\,$\times$\,10$^{32}$ erg s$^{-1}$. \citet{baglio16aa} have also recently reported on near-IR, optical and UV observations of 1RXS J180408.9-342058, which seem to indicate a jet break at $\sim$5$\times$10$^{14}$~Hz during the hard state of the source (see their Fig.~4).  

Theoretically, the frequency of the spectral break is expected to depend on parameters such as the luminosity, the magnetic field and/or the radius of the acceleration zone in the inner regions of the jet. \citet{russell13mnras} studied the dependence of the spectral break on these parameters in twelve BH XRBs and concluded that there was no obvious scaling of the break frequency with luminosity. Shortly after, observations of the BH MAXI~J1836-194 confirmed this picture \citep{1836:russell14mnras}. Currently, the multi-frequency observations of a given source as it evolves throughout an outburst have led to the picture that the jet break frequency is driven primarily by the changing structure of the accretion flow, rather than by luminosity \citep{1659:horst13mnras,gx339:corbel13bmnras,1836:russell14mnras}, at least in transitional states. In particular, the frequency at which the break occurs could be related to the offset distance from the central BH at which the acceleration occurs, and has been found to lie in the range of 10-1000 r$_g$ from the central BH for brighter hard states \citep{1836:russell14mnras}. If so, transitional states should be showing breaks at lower frequencies. Interestingly, this is precisely what is found by \citet{koljonen15apj}, who studied BHs between 10 and 10$^{9}$ M$_{\odot}$ and found an empirical correlation between the frequency of the spectral break and the index of the power law at X-ray energies during times when the jets are produced, indicating that the internal properties of the jet rely on the conditions of the plasma close to the compact object (and not, e.g., on the mass of the BH). 

In this paper, we determined the spectral break for \srcone\ to lie between 0.13 and 1.1~$\times$10$^{14}$~Hz, with a most likely frequency of 6.5$\times$10$^{13}$~Hz, slightly higher than (but consistent within the errors with) 4U~0614+091. Since 4U~0614+091 was in a hard state, we could expect the break for \srcone\ to lie at lower frequencies due to its possible transitional state. Given that we do not expect any dependence with mass for NSs, we are left with the possibility that the magnetic field may also be playing an important role. While the magnetic field has not been measured for either 4U~0614+091 or \srcone, both sources are expected to have low magnetic fields. For NSs, the next step is observing a given source at different epochs to establish if the break does move following the same parameters as for BHs. We expect ALMA observations to be crucial for these studies due to its high sensitivity at mm/sub-mm frequencies, allowing detections of weak NSs such as \srctwo\ at high significance in a few hours. 

\subsection{Additional sources detected with ALMA}

We detect one and two additional sources in the fields of view of \srcone\ and \srctwo, respectively (see Fig.~\ref{fig:image1728}).

\begin{figure*}[ht]
\includegraphics[angle=0.0,width=0.32\textheight]{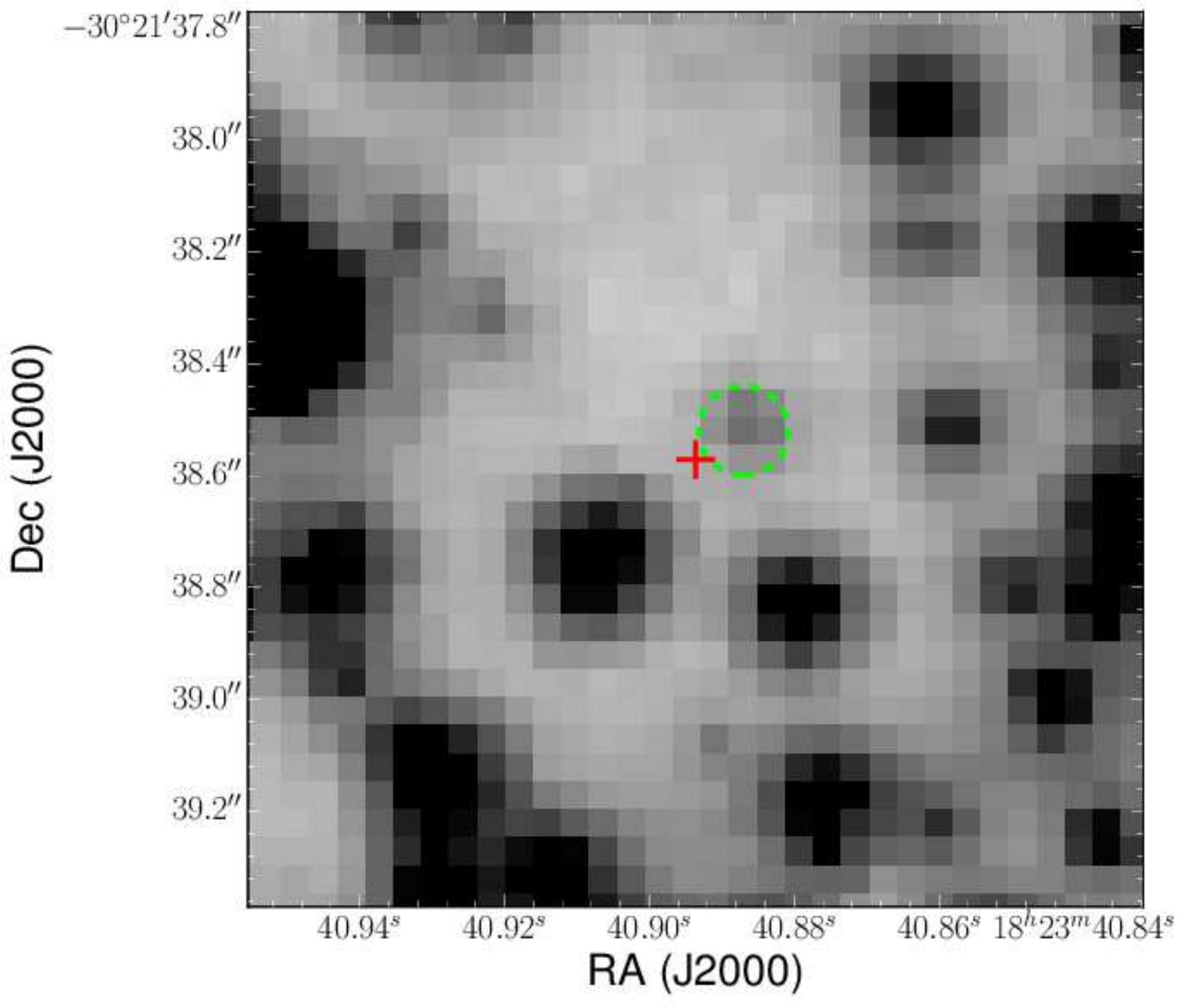}
\includegraphics[angle=0.0,width=0.32\textheight]{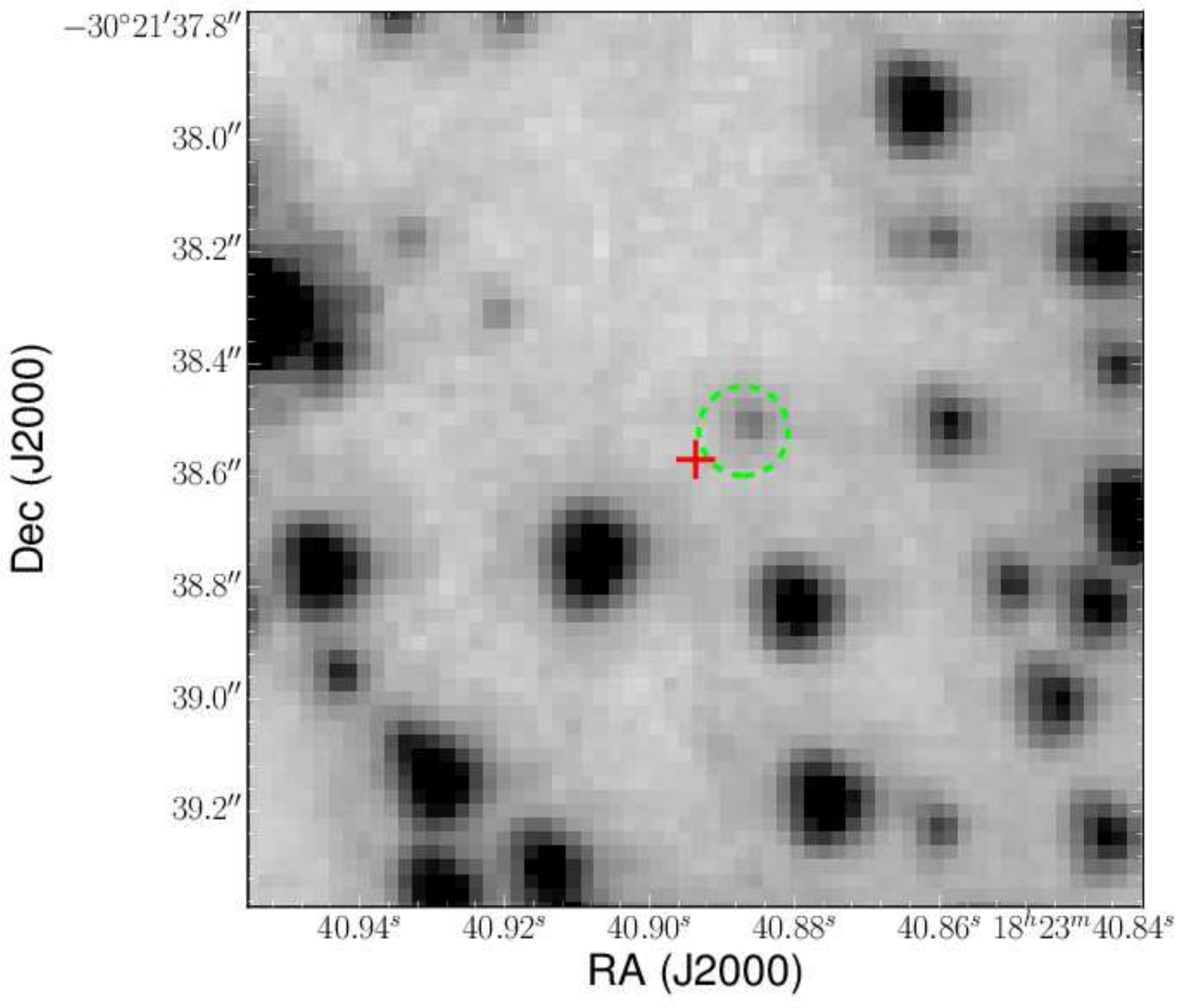}
\includegraphics[angle=0.0,width=0.33\textheight]{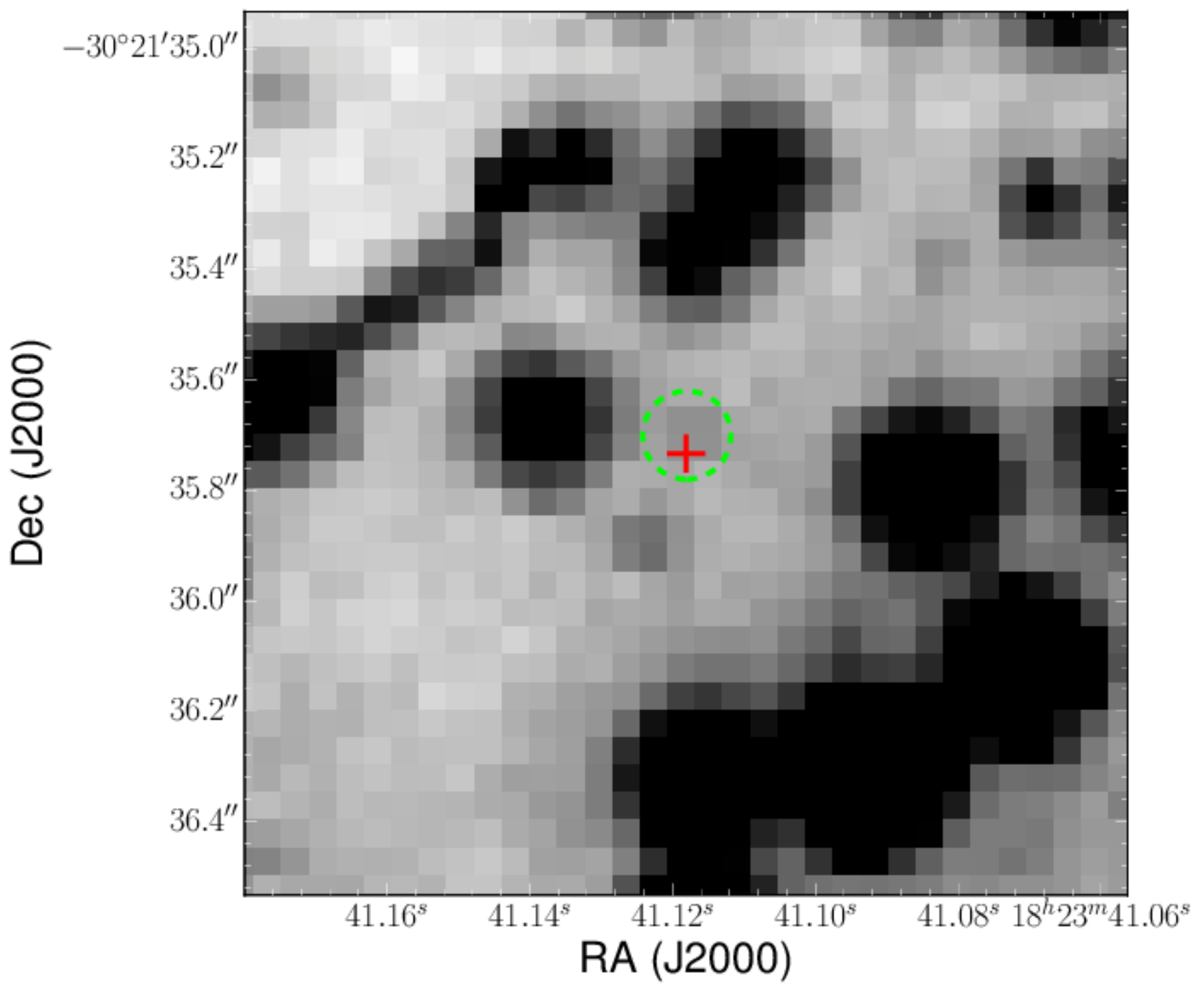}
\hspace{2.cm}
\includegraphics[angle=0.0,width=0.33\textheight]{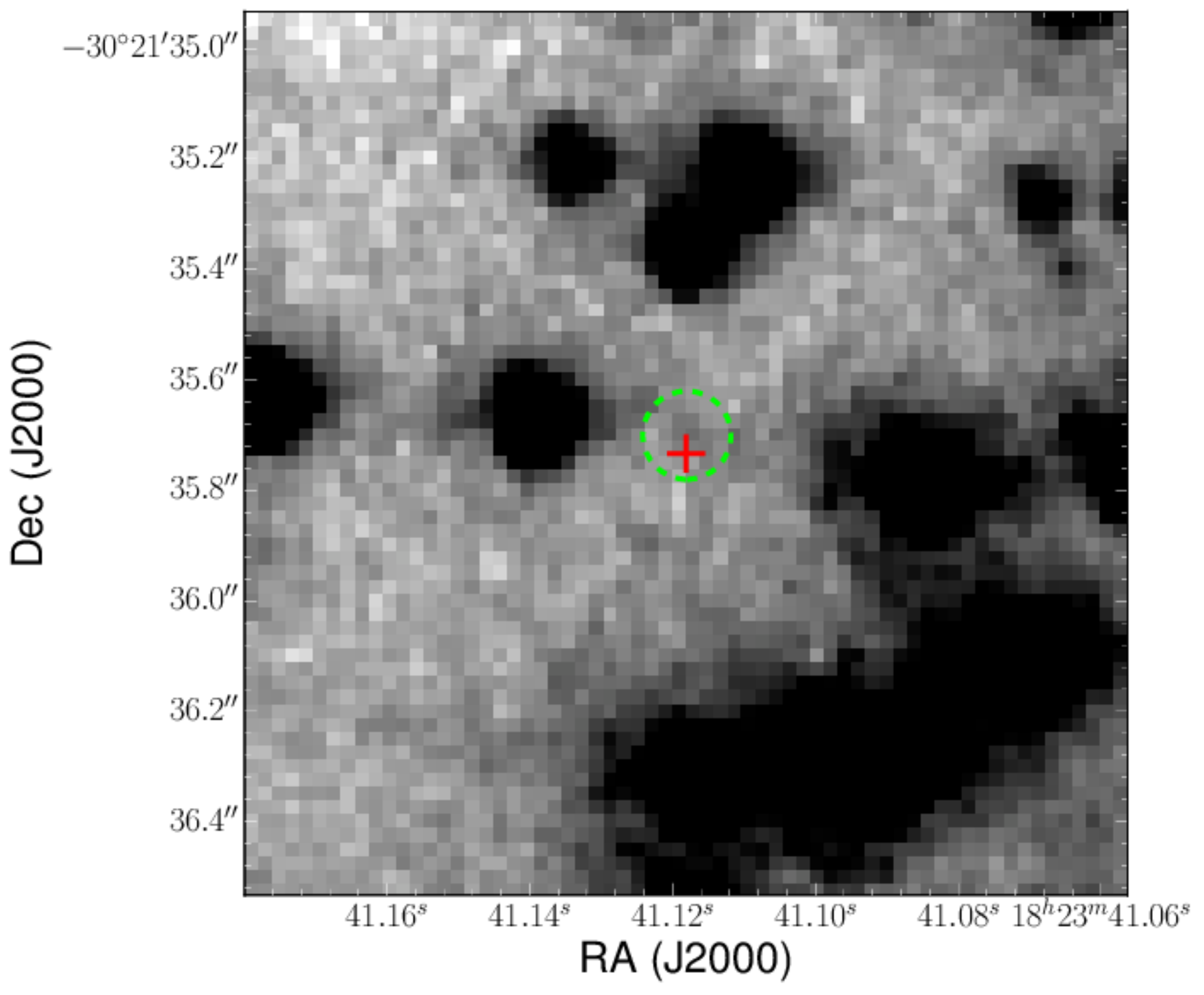}
\caption{HST images with red, F814W, filter (left panels) and blue, F435W, filter (right panels) of the sources ALMA J182340.8936--302138.571 (upper panels) and ALMA J182341.1181--302135.733 (lower panels) in NGC~6624. The red crosses mark the position of the ALMA sources and the green circles the 1\,$\sigma$ position error (0.08\arcsec, see text) of the potential counterparts.}
\label{fig:hst}
\end{figure*}

We searched for possible optical counterparts to the additional sources in NGC~6624 in {\it Hubble Space Telescope} (HST) images taken with red (F814W) and blue (F435W) filters. We re-calibrated the astrometry of the HST data using stars from the Two Micron All Sky Survey catalog \citep[2MASS; ][]{skrutskie06aj}, obtaining an astrometric error of 0.08\arcsec (1\,$\sigma$). We found a possible optical counterpart for each of the two ALMA sources (see Fig.~\ref{fig:hst}), with de-reddened \citep[E(B-V) = 0.28; ][]{valenti04mnras} Vega-system magnitudes of B\,=\,22.75\,$\pm$\,0.06, I\,=\,20.77\,$\pm$\,0.07 (for the counterpart of ALMA J182340.8936--302138.571) and B\,=\,25.58\,$\pm$\,0.22, I\,=\,22.78\,$\pm$\,0.18 (for the counterpart of ALMA J182341.1181--302135.733). Using the Dartmouth Stellar Evolution Program \citep{dotter08apjs}, assuming an age of 10.6 Gyr, a metallicity [Fe/H]\,=\,--0.69, enhancement [$\alpha$/Fe]\,=\,0.4, and distance modulus [M-m]$_0$\,=\,14.63 \citep{valenti04mnras,valenti11mnras}, we found that these counterparts correspond to main-sequence stars with masses of $\sim$0.63\,M$_{\odot}$ and $\sim$0.43\,M$_{\odot}$.

Emission from main-sequence stars at mm frequencies has  only been detected recently \citep{liseau15aa}. They reported a flux density of 10 -- 30~mJy beam$^{-1}$ at a distance of 1.35~pc, i.e. orders of magnitude lower than the flux densities from the sources detected in this paper after correcting for the distance to NGC~6624, thus ruling out the possibility that the mm emission comes from the optical 
counterparts if they are associated to the globular cluster (see below). Another possibility is that the mm emission originates in a compact object that belongs to a binary with a main sequence star. If the compact object were an accreting neutron star or stellar-mass black hole, it is strange that such sources (at distances of 5\arcsec\ and 9\arcsec\ from \srctwo) were never found in existing radio images of the cluster. Conversely, it is not so surprising that such sources were never observed in X-rays, since only $\chandra$ has sufficient angular resolution to resolve them, but \srctwo\ is so bright that it prevents the detection of any nearby source. A possibility is that they were caught in outburst during our ALMA observation, although we consider it unlikely that both of them went into outburst at the same time. Alternatively, if the compact object were a white dwarf, we could easily explain the sources being missed in previous radio or X-ray images due to their lower luminosity. However, a comparison with nearby cataclysmic variables (CV) is again problematic. CVs have never been detected at mm frequencies. At radio frequencies,  the few existing detections have flux densities of $\approxlt$0.3~mJy beam$^{-1}$ for sources at $\approxlt$ 300~pc \citep{coppejans15mnras,coppejans16mnras}, thus implying much lower luminosities than the sources reported here.  Similarly, mm emission from dust discs from stars \citep{klein03apj} or magnetic dwarfs \citep{williams15apj} seems unlikely due to their expected faintness. A further complication is that ALMA J182340.892--302138.56 is resolved with respect to the synthesized beam of 0.34\arcsec$\times$0.32\arcsec. This implies an extension of the mm emission of $\approxgt$10$^{16}$~cm. Such an extended emission is difficult to explain even for compact jets of X-ray binaries. So far, emission from a compact jet has only been resolved in a handful of cases and determined to be $\approxlt$10$^{15}$~cm \citep[e.g.][]{cygx1:stirling01mnras}. In conclusion, any association with sources in the globular cluster looks problematic. 

Interestingly, no sources were detected in an ALMA observation of the globular cluster 47~Tuc at 230~GHz \citep{mcdonald15mnras}. However, those observations have a significantly lower continuum  sensitivity since they were targeted at detecting CO lines and consequently had very narrow bandwidth. In the public calibrated products available from the ALMA archive for these observations we found a continuum noise of 0.2~mJy~beam$^{-1}$ for a beam size of 2 arcseconds, i.e. the noise is ~10 times higher than in our images. This noise level would have resulted in a detection of only the brightest source in both of our images (i.e. one source per image). 

Given the high number density of stars from the cluster NGC~6624 within the ALMA field of view, the two ALMA sources and their proposed stellar counterparts might actually not be associated. We estimate that 1--2 stars could fall by chance within the 2\,$\sigma$ error circles (dominated by the astrometric uncertainty of the HST images) of the ALMA sources. This suggests that the two previously discussed stars might not be the sources of mm emission. Alternatively, the sources could be background AGN or star forming galaxies. 
However, if they were AGN we should have detected them in the radio images assuming that their spectrum is flat from radio to mm frequencies. Regarding star forming galaxies, \citet{oteo16apj} estimated a number density of 1.7\,($\pm$\,1.4) $\times$10$^{4}$ background sources per square degree at 345~GHz with flux densities $>$~0.4~mJy beam$^{-1}$. This implies 0--3 sources in the field of view of our ALMA observations, consistent with our number of detections. 

\section{Conclusions}

We detected the two NS atoll sources \srcone\ and \srctwo\ at mm frequencies for the first time. \srcone\ was most likely transitioning from a hard to a soft state and its radio to IR  SED is consistent with emission from a jet with a break from optically thick to optically thin emission at $\sim$6$\times$10$^{13}$~Hz, at a slightly higher frequency than (but consistent within the errors with) that detected for 4U~0614+091 during a hard state. Given that low-magnetic field NSs are expected to have similar magnetic field values and masses, observations of more spectral breaks as a function of spectral state could prove crucial to unveil the key ingredients of jet evolution. 

The observations of \srctwo\ confirm the existence of significant emission in the radio and mm bands during the soft state for NSs, as opposed to the quenching of such emission in BHs. Its SED might indicate variability among the radio and mm observations, emphasizing the need for strictly simultaneous observations at different frequencies for these studies. Further observations with ALMA at low frequencies ($\sim$ 100~GHz) simultaneous to observations with ATCA at frequencies of 9~GHz and above could shed some light on the origin of this SED. 


\begin{acknowledgements} 
This paper makes use of the following ALMA data:
   ADS/JAO.ALMA\#2013.1.01013.S. ALMA is a partnership of ESO (representing
   its member states), NSF (USA) and NINS (Japan), together with NRC
   (Canada) and NSC and ASIAA (Taiwan), in cooperation with the Republic of
   Chile. The Joint ALMA Observatory is operated by ESO, AUI/NRAO and NAOJ.
   The Australia Telescope Compact Array is part of the Australia Telescope National Facility which is funded by the Australian Government for operation as a National Facility managed by CSIRO. JCAMJ is the recipient of an Australian Research Council Future Fellowship (FT140101082). S.M. acknowledges support by the Spanish Ministerio de Econom\'ia y Competitividad (MINECO) under grants AYA2013- 47447-C3-1-P, MDM-2014-0369 of ICCUB (Unidad de Excelencia ÔMar\'ia de MaeztuÕ). This research made use of APLpy, an open-source plotting package for Python hosted at http://aplpy.github.com.
\end{acknowledgements}


\bibliographystyle{aa}
\bibliography{alma}

\end{document}